\documentstyle[11 pt]{article}
\newcommand{\be}{\begin{equation}}
\newcommand{\ee}{\end{equation}}
\newcommand{\bea}{\begin{eqnarray}}
\newcommand{\eea}{\end{eqnarray}}

\begin{document}
\vspace{1.0in}

\begin{center}
{\bf Dynamics of (SUSY) AdS Space Isometry Breaking\footnote{to appear in Proceedings of IRGAC 2006: 2$^{nd}$ International Conference on Quantum Theories and Renormalization Group in Gravity and Cosmology, Barcelona}}
\end{center}

\begin{center} 
S.T. Love\footnote{e-mail address: loves@physics.purdue.edu} \\
{\it Department of Physics\\ 
Purdue University\\
West Lafayette, IN 47907-2306}
~\\
\end{center}
\vspace{2pt}

\small

\begin{abstract}
\noindent
{\scriptsize Actions governing the dynamics of the Nambu-Goldstone modes resulting from the spontaneous breaking of the  
$SO(4,2)$ and $SU(2,2|1)$ isometries of five dimensional anti-de Sitter space ($AdS_{5}$) and SUSY $AdS_{5}\times S_1$ spaces respectively due to a restriction of the motion to embedded four dimensional $AdS_{4}$ space and  four dimensional Minkowski space ($M_4$) probe branes are presented. The dilatonic Nambu-Goldstone mode governing the motion of the $M_4$ space probe brane into the covolume of  the SUSY $AdS_5\times S_1$ space is found to be unstable. No such instablility appears in the other cases. Gauging these symmetries leads to an Einstein-Hilbert action containing, in addition to the  gravitational vierbein, a massive Abelian vector field coupled to gravity. 
}

\end{abstract}
\vspace{2pt}

\small

Conformal and superconformal invariance play a pivotal role in many currently investigated theoretical models. A major advance which has further ellucidated these studies was the conjectured correspondence between certain (super) conformal field theories and theories formulated on anti-de Sitter ($AdS$) and supersymmetric (SUSY) $AdS$ spaces\cite{review}. Here I present various dynamical consequences for field theories on $AdS$ spaces and their supersymmetric extensions which arise due to the spontaneous breakdown of some of their space-time symmetries when the motion is restricted  to  lower dimensional probe branes.

A background $AdS_{5}$ space is characterized by a constant Ricci scalar curvature, $R=-20m^2$, and has the isometry group $SO(4,2)$ whose generators, $M^{MN}=-M^{NM}, ~ M,  N = 0,1,2,3,4,5$ satisfy the algebra: 
\be
[M_{MN}, M_{LR}]= i(\hat{\eta}_{ML}M_{NR}-\hat{\eta}_{MR}M_{NL}-\hat{\eta}_{NL}M_{MR}+\hat{\eta}_{NR}M_{NL}) 
\ee
where $\hat{\eta}_{MN}$ is a diagonal metric tensor with signature $(-1,+1,+1,+1,+1,-1)$. For embedding $AdS_4$ space, it is useful to parametrize the $AdS_5$ space with the coordinates $\rho, x^\mu ~;~ \mu = 0,1,2,3$, so that the $AdS_{5}$ space invariant interval takes the form
\bea
ds^2 &=& e^{2A(\rho)} \eta_{\lambda\mu} \bar{e}_\nu^{~\sigma}(x)\bar{e}_\sigma^{~\lambda}(x) dx^\mu dx^\nu  +(d\rho)^2 
\label{AdSIV}
\eea
where $A(\rho)=\ell n [cosh(m\rho)]$ is a warp factor and 
$\bar{e}_\mu~^\nu(x)=\frac{sinh(\sqrt{m^2x^2}}{\sqrt{m^2x^2}}P_{\perp\mu}^\nu(x)+P_{\parallel\mu}^\nu(x)$ is the $AdS_4$ vierbein. Here $P_{\perp~\mu}~^\nu(x)=\eta_\mu^\nu -\frac{x_\mu x^\nu}{x^2}$ and $P_{\parallel~\mu}~^\nu(x)=\frac{x_\mu x^\nu}{x^2}$ are tranverse and longitudinal projectors respectively while $\eta^{\mu\nu}$ is the $4d$ Minkowski space metric tensor with signature $(-1,1,1,1)$. The $AdS_4$ invariant interval is obtained from the 
$AdS_{5}$  interval by setting $\rho =0$ which in turn gives the orientation of $AdS_4$ brane in $AdS_{5}$ space. 
Such an embedding of $AdS_{4}$ space spontaneously breaks the isometry group of the $AdS_{5}$ space from $SO(4,2)$ to $SO(3,2)$. Introducing the (pseudo-) translation generators defined as ${P}_\mu = m{M}_{5 \mu}~;~D=mM_{54}$, the broken generators are then identified as $D $ and $M_{4\mu }$.  

A model independent way of encapsulating the long wavelength dynamical constraints imposed by this spontaneous symmetry breakdown is to realize the $SO(4,2)$ isometry nonlinearly  on the Nambu-Goldstone boson fields $\phi$ and $v^\mu$ associated with the broken symmetry generators $D$  and $M_{4 \mu }$ respectively.  Using coset methods\cite{CCWZ}, the $AdS_5$ vierbein factorizes\cite{CLNtV1} as $e_\mu~^\nu= \bar{e}_\mu~^\lambda N_\lambda~^\nu $ where $\bar{e}_\mu^{~\nu}$ is the $AdS_4$ vierbein  and $N_\lambda~^\nu = cosh(m\phi)\biggr\{ [P_{\perp~\lambda}~^\nu (v) +cos (\sqrt{v^2}) P_{\parallel~\lambda}~^\nu(v)] +{ D}_\lambda \phi \frac{sin(\sqrt{v^2})}{\sqrt{v^2}} v^\nu \biggr\}$ with $D_\mu =\frac{1}{cosh(m\phi)}\bar{e}_\mu^{-1~\nu}\partial_\nu$  the $AdS_4$ covariant derivative. The resultant $SO(4,2)$ invariant action is $S=-\sigma \int d^4x ~det~e$ with $\sigma$ the $AdS_4$ brane tension. Since this action is independent of $\partial_\mu v_\nu$, the Nambu-Goldstone field  $v_\mu$ is nondynamical\cite{IO} and it can be eliminated using  its field equation  $v^\mu \frac{tan(\sqrt{v^2})}{\sqrt{v^2}}=D^\mu\phi$ so that  the action can be recast as\cite{CLNtV1}
\bea
S&=& -\sigma \int d^4x ~det~\bar{e}~ cosh^4(m\phi) \sqrt{1+ D_\mu\phi D^\mu\phi}
\label{adsa1}
\eea
This Nambu-Goldstone field action contains a mass term, $m^2_\phi =  4m^2 $, along with non-derivative interactions and constitutes an 
$AdS$ generalization of Nambu-Goto action\cite{ng}. Using the factorized  $AdS_{5}$ vierbein, along with $v$ field equation, the invariant interval for $AdS_{5}$  space takes the form
\bea
ds^2 &=&  e^{2A(\phi)} \eta_{\lambda\mu} \bar{e}_\nu^{~\sigma}(x)\bar{e}_\sigma^{~\lambda}(x)dx^\mu dx^\nu +(d\phi(x))^2 
\eea
with $A(\phi)=\ell n[cosh(m\phi)]$. This has the same structure as the invariant interval of $AdS_{5}$ space, Eq. (\ref{AdSIV}),  obtained previously after the identification of $\phi(x)$ with the covolume coordinate $\rho$. As such, $\phi(x)$ describes the motion of $AdS_4$ brane into remainder of $AdS_{5}$ space.

Next, we embed a $4$-dimensional Minkowski space, $M_4$, probe brane into $AdS_{5}$ space\cite{CL06}. For such an embedding, it proves convenient to introduce a different set of $AdS_5$ coordinates $x^\mu, x^4$ 
so that the $AdS_{5}$ invariant interval is 
\be
ds^2= e^{2mx_4}dx^\mu \eta_{\mu\nu}dx^\nu +(dx_4)^2
\label{inmb}
\ee 
which reduces to the $M_4$ space invariant interval at $x_4=0$. Thus inserting a Minkowski space probe brane at $x_4=0$, the broken generators are identified as  $D ~,~ {M}^{4\mu }$ and the $SO(4,2)$ isometry can be nonlinearly realized on the Nambu-Goldstone bosons, the  dilaton, $\phi$, and $v^\mu$ associated with these broken symmetry generators. The coset space construction allows the  extraction the $AdS_5$ vierbein as ${e}_\mu^{~\nu}= e^\phi [P_{\perp\mu}^\nu(v) +P_{\parallel\mu}^\nu(v)cos(\sqrt{m^2v^2})]-\partial_\mu\phi v^\nu \frac{sin(\sqrt{m^2v^2})}{\sqrt{m^2v^2}}$. 
Once again, $v^\mu$ is not independent dynamical degree of freedom.  Eliminating it using its field equation $v^\mu \frac{tan(\sqrt{v^2})}{\sqrt{v^2}}=-e^{\phi}\partial^\mu \phi$ yields the invariant action term $-\sigma \int d^4x e^{4\phi } \sqrt{1+\frac{1}{m^2}e^{-2\phi}\partial_\mu\phi\eta^{\mu\nu}\partial_\nu\phi}$ 
while the invariant interval can be written as
\bea
ds^2 =e^{2\phi} dx^\mu\eta_{\mu\nu}dx^\nu +\frac{1}{m^2}(d\phi)^2 
\label{mbin}
\eea
This has the same form as $AdS_{5}$ invariant interval of Eq. (\ref{inmb}) after the identification of $\phi\Leftrightarrow\frac{1}{m}x_4$. Thus the dilaton dynamics describes the motion of brane into the covolume of $AdS_{5}$ space. In the above, a particular combination for the broken generators was chosen.
An alternate, equally valid, choice is $D$ and 
$K^\mu = \frac{1}{m^2}{P}^\mu -\frac{1}{m}{M}^{4\mu }$. 
This, in turn, leads to the $4$-dimensional conformal algebra. Moreover, since the generators $K^\mu$ and ${M}^{4\mu }$ differ only by the unbroken translation generator $P^\mu$, the action  is also invariant under $4$-dimensional conformal transformations. Since $e^{4\phi }$ transforms as total divergence under conformal transformations, the invariant term $ \int d^4x e^{4\phi }$ can be subtracted producing the $SO(4,2)$ invariant action\cite{F}-\cite{GKW}
\bea
S&=& -\sigma \int d^4x e^{4\phi } [\sqrt{1+\frac{1}{m^2}e^{-2\phi}\partial_\mu\phi\partial^\mu\phi}-1] 
\label{mba1}
\eea
which is defined so as to have zero vacuum energy.

Now consider embedding $M_4$ and $AdS_4$ branes in SUSY $AdS_5 \times S_1$ space\cite{Zumino}-\cite{P}. The supersymmetric $AdS_5 \times S_1$ isometry algebra, $SU(2,2|1)$, includes the generators $M^{\mu\nu}, P^\mu, M^{4\mu}, D$ of the $SO(4,2)$ isometry algebra, the 
SUSY fermionic charges $Q_\alpha, \bar{Q}_{\dot\alpha}, S_\alpha, \bar{S}_{\dot\alpha}$  
and the $R$ charge which is the  generator of the $U(1)$ isometry of $S_1$. Embedding an $M_4$ probe brane at $x^4 =0$ breaks the space-time symmetries generated by ${P}_4$ and ${M}_{4\mu}$, as well as all the supersymmetries  and the $R$ symmetry. This $SU(2,2|1)$ isometry algebra of the super-$AdS_5 \times S_1$ space can be nonlinearly realized on the  Nambu-Goldstone modes of the broken symmetries\cite{CL06}. These are the dilaton, $\phi$, and $v^\mu$ associated with $D$ and $M_{4\mu }$ respectively, the Goldstinos $ \lambda_\alpha~,~ \bar{\lambda}_{\dot\alpha}$ and $\lambda_{S\alpha}~,~\bar\lambda_{S\dot\alpha}$ of the spontaneously broken supersymmetries, $Q_\alpha, \bar{Q}_{\dot\alpha}, S_\alpha, \bar{S}_{\dot\alpha},$ and the $R$-axion $a$. The Nambu-Goldstone bosonic modes $v^\mu$ and the Goldstinos $\lambda_{S\alpha}~,~\bar\lambda_{S\dot\alpha}$  are not independent dynamical degrees of freedom\cite{IO} but rather are given in terms of the dilaton and Goldstinos $ \lambda_\alpha~,~ \bar{\lambda}_{\dot\alpha}$ as
$ v^\mu = \partial_\mu \phi +...$, 
$\lambda_{S\alpha}= (\sigma^\mu\partial_\mu\bar\lambda)_\alpha +...$ and 
$\bar{\lambda}_{\bar{S}\dot\alpha}=(\partial_\mu\lambda \sigma^\mu)_{\dot\alpha}+...$
After elimination of the non-dynamical Nambu-Goldstone modes, the resultant invariant action is   
\bea
S=-\sigma \int d^4x~ e^{4\phi} ~det~A~\sqrt{1+\frac{e^{-2\phi}}{m^2}{\cal D}_\mu\phi {\cal D}^\mu\phi}[1+e^{-2\phi} {\cal D}_\mu a {\cal D}^\mu a][1+B]
\label{GS}
\eea
where $A_\mu^{~\nu}= \eta_\mu^\nu +i (\lambda \stackrel{\leftrightarrow}{\partial}_\mu\sigma^\nu\bar\lambda)$ 
is the Akulov-Volkov vierbein\cite{AV}, ${\cal D}_\mu= A^{-1 \nu}_\mu \partial_\nu $
is the SUSY covariant derivative and $B$ is a somewhat lengthy sum of terms all of which are least bilinear in the Goldstino fields and contain at least two derivatives\cite{CL06}.  The action is an invariant synthesis of Akulov-Volkov and Nambu-Goto actions. Note that the pure dilatonic part of the action (obtained by setting the Goldstinos and $R$-axion to zero so that $A_\mu^{~\nu}=\delta_\mu^\nu$ and $B=0$) reproduces the previous action  of the Minkowski space $M_4$ probe brane in $AdS_{5}$ without SUSY. As such, the dilaton $\phi$ describes the motion of the probe brane into the rest of the $AdS_5$ space. However, in this case, because of the spontaneous breakdown of the complete SUSY, there is no invariant that can be added to the action to cancel the vacuum energy such as one was able to achieve in the non-supersymmetric Minkowski space probe brane case (c.f. Eq. (\ref{mba1})). It follows that the dilaton feels an $e^{4\phi}$ potential which, in turn,  contains a destabilizing term linear in $\phi$ driving the dilaton field to $\phi \rightarrow -\infty$. Since the dilaton describes the motion of the probe Minkowski $M_4$ brane into the remainder of $AdS_5$ space, it follows that the SUSY $AdS_5$ space cannot sustain the Minkowski space brane.  

The alternate combination of broken generators $D$ and $K_\mu = \frac{1}{m^2}({P}_\mu -2m M_{4  \mu })$ can also be defined. This leads to the 4-d superconformal algebra. The spontaneously broken symmetries are $R$, dilatations ($D$), special conformal ($K^\mu$), SUSY ($Q_\alpha~,~\bar{Q}_{\dot\alpha}$) and SUSY conformal ($S_\alpha~,~\bar{S}_{\dot\alpha}$). Since the generators $K^\mu$ and $M^{\mu 4}$ differ only by unbroken translation generator $P^\mu$, the action ($\ref{GS}$) is invariant under superconformal transformations. Once again the potential for the dilaton $\phi$ is unstable and there is an incompatibility of simultaneous nonlinear realizations of SUSY and scale symmetry in four dimensional Minkowski space\cite{CLtheorem}. Alternatively expressed, the spectrum of four dimensional Minkowski space cannot include both the Goldstino and the dilaton as Nambu-Goldstone modes. Note that the origin of this unusual behavior is not simply a consequence of the introduction of a scale due the spontaneously broken SUSY. It has been shown that there is no incompatibility in securing simultaneous nonlinear realization of spontaneously broken scale and chiral symmetries\cite{B} where a scale is also introduced. In that case, the spectrum of the effective Lagrangian admits both pions and a dilaton.

On the other hand, the invariant action for the dilaton $\phi$ and Goldstinos obtained by embedding an $AdS_4$ probe brane in SUSY $AdS_5\times S_1$ space has, in addition to other modifications, an overall prefactor of $cosh^4(m\phi)$ instead of $e^{4\phi}$. Thus, in this case, there is no destabilizing linear in $\phi$ term. Consequently an $AdS_4$ brane can be embedded in SUSY $AdS_5\times S_1$ space and the spectrum can admit both a massive dilaton and massive Goldstinos.

Thus far, we have focused on a fixed background $AdS_{5}$ space and the actions constructed are invariant under a nonlinear realization of the global isometry group $SO(4,2)$.  In order to describe the dynamics of an oscillating brane embedded in curved space, we need to have invariance under local $SO(4,2)$ transformations and additional gauge fields including dynamical gravity must be introduced. The dynamics of the brane embedded in curved space is then described by a brane localized massless graviton\cite{KR}-\cite{S} represented by a dynamical metric tensor $g_{\mu\nu}$ and a vector field 
$A_\mu (x)$. As a consequence of the Higgs mechanism, the vector field is massive\cite{Por}.  The action for these fields is once again derived in a model independent manner using coset methods. Isolating the physical degrees of freedom by working in unitary gauge defined by setting $\phi=0$ and $v^a=0$, the action takes the form\cite{CLNtV2}
\bea
&&S=\int d^4x ~\sqrt{-det~g}~\{-\frac{1}{16\pi G_N}(2\Lambda + R) -\frac{1}{4}F_{\mu\nu}g^{\mu\rho}g^{\nu\sigma}F_{\rho\sigma}\cr
&&~~~~~~~~~~+\frac{1}{2}A_\mu[(M^2 +c_1 R)g^{\mu\nu} +c_2 R^{\mu\nu}]A_\nu\}
\eea
where $\Lambda$ is the cosmological constant, $G_N$ is Newton's constant, $R^{\mu\nu} (R)$ is the full (background plus dynamical) Ricci tensor (scalar), while $F_{\mu\nu}=\partial_\mu A_\nu-\partial_\nu A_\mu$ is the Abelian field strength and $c_1, c_2$ are constants. This is recognized as the action of a massive Proca field $A_\mu$ with independent mass parameter $M$ interacting with either $AdS_4$ or $M_4$ Einstein gravity. When coupled to the Standard Model, this Abelian vector field transforms analogously to the weak hypercharge gauge field and thus will lead to a $Z^\prime$ boson in the spectrum. Note that since the vector mass $M$ is an independent parameter, it is nonzero even in the flat space limit ($m=0$) and consequently such a massive Abelian Proca field also appears when an $M_4$ brane probe is inserted in $M_5$ space in a locally invariant manner.\\

\noindent This was supported in part by the U.S. Department of Energy under grant DE-FG02-91ER40681 (Task B). I thank T.E. Clark for an enjoyable collaboration.

\end{document}